\documentclass[pra,letterpaper,twocolumn,superscriptaddress,floatfix]{revtex4}

\usepackage{graphicx,psfrag,amsmath,amssymb,amsfonts,bbm,latexsym,color,dcolumn,bm}

\begin{document}

\title{Concepts for a deuterium-deuterium fusion reactor}

\author{Roberto Onofrio}

\affiliation{\mbox{Dipartimento di Fisica e Astronomia ``Galileo Galilei'', Universit\`a  di Padova, 
Via Marzolo 8, Padova 35131, Italy}}

\affiliation{\mbox{Department of Physics and Astronomy, Dartmouth College, 6127 Wilder Laboratory, 
Hanover, NH 03755, USA}}

\begin{abstract}
We revisit the assumption that reactors based on deuterium-deuterium (D-D) fusion processes
have to be necessarily developed after the successful completion of experiments and
demonstrations for deuterium-tritium (D-T) fusion reactors. Two possible mechanisms for enhancing
the reactivity are discussed. Hard tails in the energy distribution of the nuclei, through
the so-called $\kappa$-distribution, allow to boost the number of energetic nuclei available
for fusion reactions. At higher temperatures than usually considered in D-T plasmas, vacuum   
polarization effects from real $e^+e^-$ and $\mu^+\mu^-$ pairs may provide further speed-up  
due to their contribution to screening of the Coulomb barrier. 
Furthermore, the energy collection system can benefit from the absence of the lithium blanket, both 
in simplicity and compactness. The usual thermal cycle can be bypassed with comparable efficiency 
levels using hadronic calorimetry and third-generation photovoltaic cells, possibly allowing to extend 
the use of fusion reactors to broader contexts, most notably maritime transport.
\end{abstract}

\maketitle

\section{Introduction}

It is usually assumed that the first commercial fusion reactors will be based on D-T mixtures,
and within this frame a well-defined path has been paved with the ongoing development of ITER, following
the successful operation of JET and other tokamaks which have produced significant amount of fusion power.
This path has two recognized drawbacks: the shortage of natural tritium sources \cite{Mitchell,Zheng}, and
the irradiation damage caused by 14.06 MeV neutrons, including the associated contamination of the reactor.
A lithium blanket to create tritium in situ is a nontrivial issue, as it must satisfy several competing
requirements, such as the need to breed tritium with easy extraction processes and generate heat, while
sustaining a large neutron flux. All these issues are avoided by using D-D reactors.
Deuterium is easily available in water, the 2.45 MeV neutrons induce a irradiation damage two orders
of magnitude smaller than the one released in D-T fusion processes, and radioactive contamination
is mitigated and mainly contained to the tritium produced in one of the two channels of the fusion reaction.
However, the D-D cross-section in the interesting energy range is about two orders of magnitude smaller than the
corresponding D-T cross-section, and therefore the requirements for igniting and self-sustaining the
reaction are more demanding \cite{Stott}.

\begin{figure*}[t]
\includegraphics[width=0.49\textwidth, clip=true]{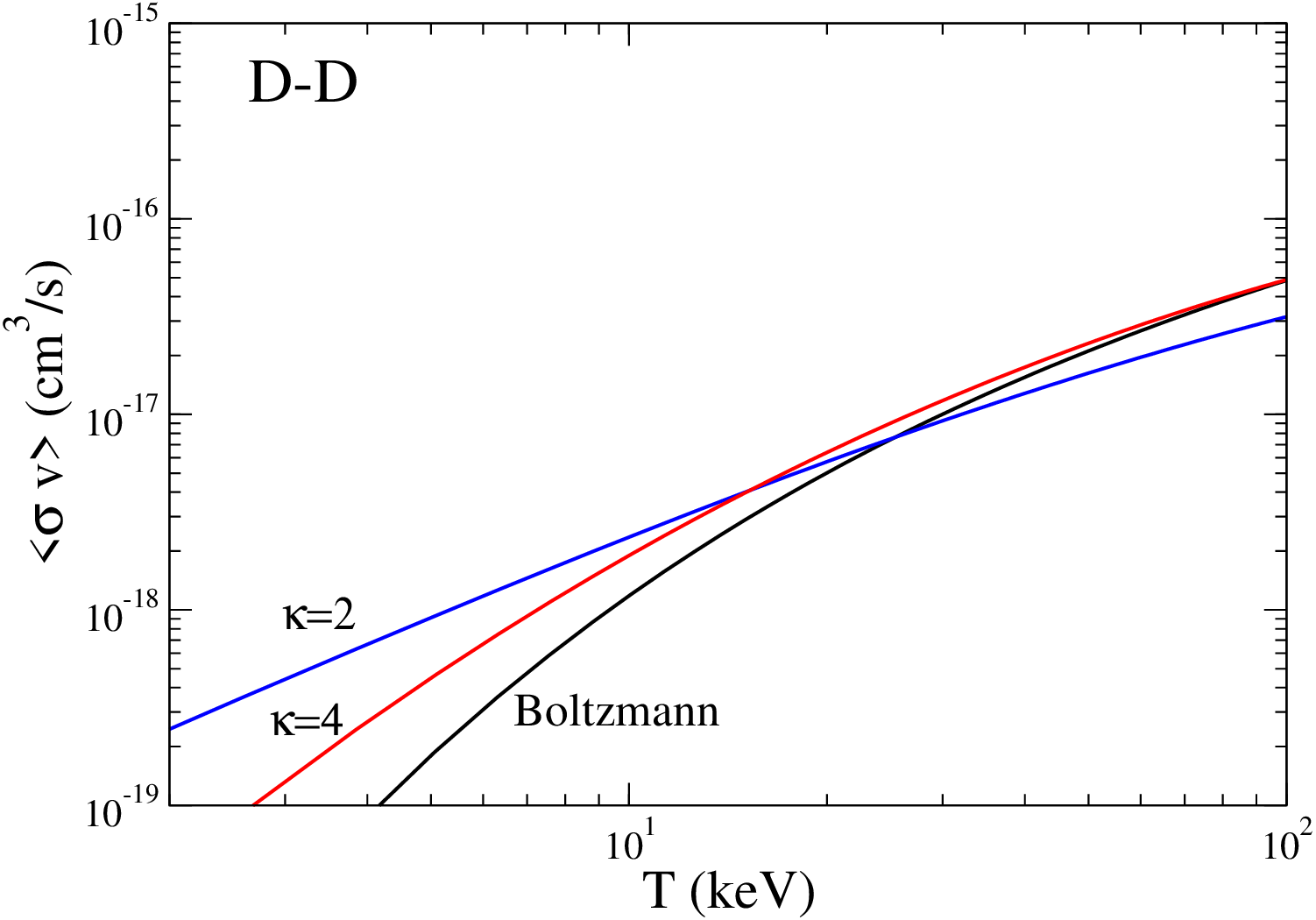}
\includegraphics[width=0.49\textwidth, clip=true]{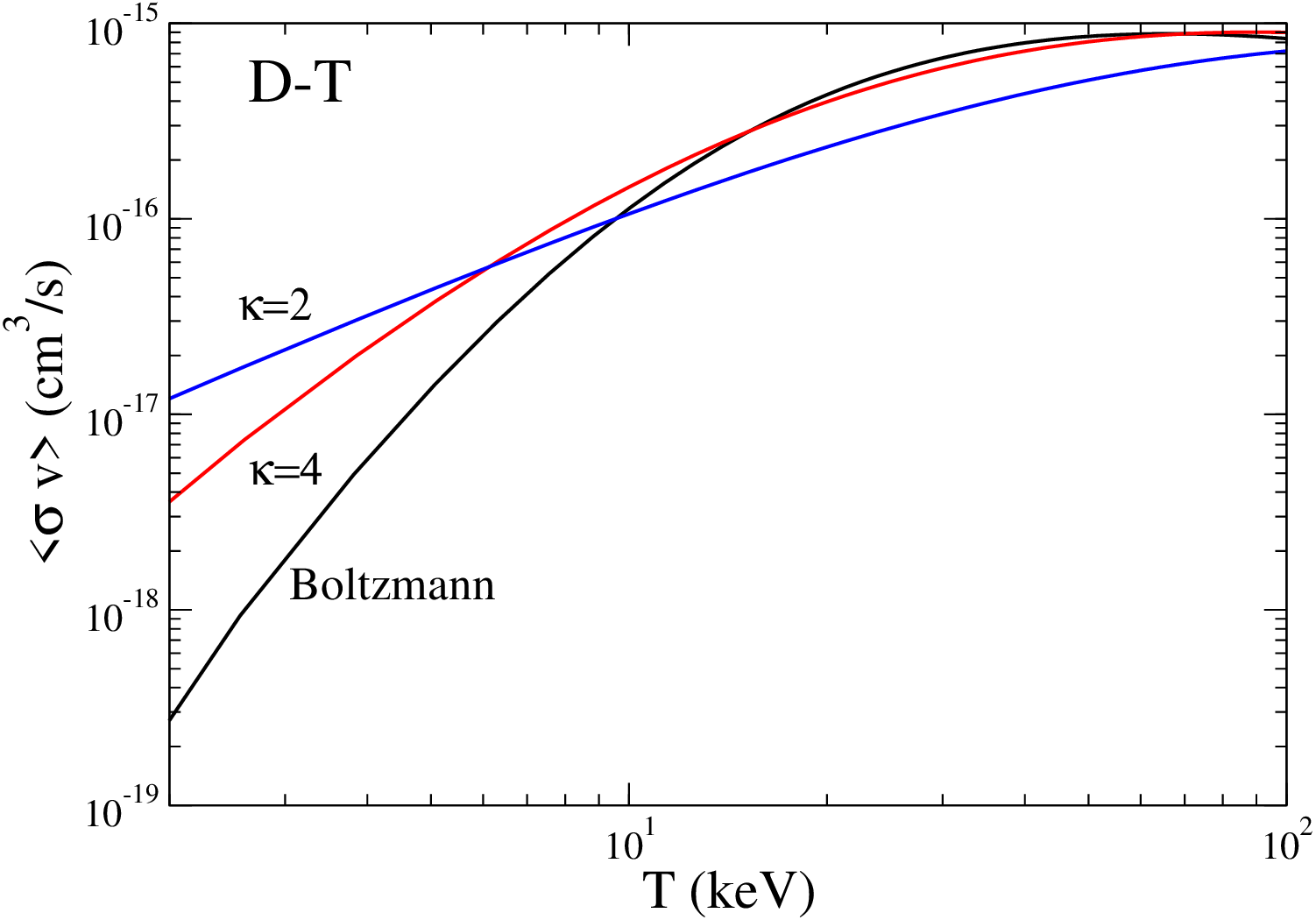}
\caption{Reactivities for D-D (left) and D-T (right) fusion processes averaged over Boltzmann energy
distributions (black) and over two $\kappa$-distributed energies, $\kappa=2$ (blue) and $\kappa=4$ (red), in
the 2-100 keV temperature range. 
Reactivities and temperatures in the two plots are expressed with the same scales allowing for an easier comparison. 
The D-D reactivity is obtained summing over both reaction channels, D(d,p)T and D(d,n)$\mathrm{{}^3He}$.}
\label{Fig1}
\end{figure*}

In this contribution, we outline two proposals for enhancing the reactivity of D-D fusion processes, and discuss the 
possibility of bypassing the thermal cycle for electricity production.
More specifically, we discuss non-Boltzmann steady state configurations represented by power-law energy 
distributions, with an estimate of the expected enhancement in the reactivity with respect to Boltzmann-state reactivities.
At higher temperatures than the one currently achieved for a D-D plasma, the possibility to create real
electron-positron pairs from vacuum may lead to a lowering of the Coulomb barrier, with a consequent
enhancement of the reactivity. Finally, we discuss the possibility for combining recent progress in
high yield scintillating materials and hadronic calorimetry with third-generation photovoltaic cells. 
This could lead to energy conversion at an initially nearly comparable efficiency with respect to a 
thermal cycle, with unknown margins for improvement, and within a compact design. 
Neutron damage induced in the calorimeter and the consequent possible decrease in the light yield 
indicate that applications of this energy conversion are limited to about 1 MW of fusion power, more 
than enough for maritime transport and for low power plants in low-density populated areas for instance.
This paper should be considered as an overview of work in progress, and as such limited by the contingency
of contributing to the Festschrift, with three distinct research directions to be pursued more quantitatively
in the close future.

\section{Enhancing the reactivity via non-Boltzmann energy distributions}

The usual comparison between D-D and D-T fusion rates relies on the hypothesis of Boltzmann energy distributions.
Considering the complex dynamics occurring for a confined plasma, it is worth to scrutinize about the adequacy of this
assumption. If the plasma is heated by neutral beam injection, a steady state situation can occur in which the heating
power is significantly higher than the relaxation rate to equilibrium of the plasma itself.
Analogous situations already occur in low-density, low-temperature plasmas characteristic, for instance,
of solar wind \cite{Vasyliunas}. Under this circumstance the high-energy tail of the distribution may
be enhanced -- basically because of the large pile-up of energy which is hardly transferred to lower
energy particles -- generating large deviations from the Boltzmann distribution.
These energy distributions, named $\kappa$-distributions, have been discussed since several decades,
see for instance \cite{Livadiotis} for a comprehensive overview with applications in astrophysical environments, and
\cite{Nicholls} for an application to solve systematic discrepancies in determining electron temperatures in
HII regions and planetary nebulae. 
The characterization of these $\kappa$-distributions requires the introduction of two parameters, the kinetic
temperature, an effective temperature such that the energy per unit of particle $U$ can still be written as
$U=3 k_B T_U/2$ as in the Boltzmann case, and the $\kappa$-parameter (with values in the $3/2 < \kappa < +\infty$ range).
The energy probability density is expressed as \cite{Nicholls}

\begin{equation}
P(E)=\frac{C(\kappa)}{(k_B T_{\mathrm{U}})^{3/2}} \frac{E^{1/2}}{\left[1+\frac{E}{(\kappa-3/2)k_B T_U}\right]^{\kappa+1}},
\end{equation}  
where $C(\kappa)=2\Gamma(\kappa+1)/[\pi^{1/2}(\kappa-3/2)^{3/2} \Gamma(\kappa-1/2)]$.

%\begin{equation}
%C(\kappa)=\frac{2}{\pi^{1/2}} \frac{\Gamma(\kappa+1)}{(\kappa-3/2)^{3/2} \Gamma(\kappa-1/2)}.
%\end{equation}

The Boltzmann distribution is recovered in the $\kappa \rightarrow +\infty$ case, with the kinetic temperature
of the $\kappa$ distribution tending to the temperature $T_{\mathrm{core}}$ of the Boltzmann distribution
interpolating its ``core'' distribution, {\it i.e.} the region of energies with the most probable population,
$T_{\mathrm{core}}=[1-3/(2\kappa)]T_{\mathrm{U}}$. The concrete value of $\kappa$ is usually determined 
from a best fit of the observed energy distribution and, as far as we are aware of, there is not yet a kinetic model
to predict its value. For now, we will assume  values of $\kappa$ as typically inferred by space
plasma physics, to have a common-sense, perhaps questionable, benchmark.

Since most of the fusion reactions occur in the high-energy tail of the energy distribution, and the
$\kappa$-distributions are characterized by a hard, power-law tails, a first exercise may consist in
evaluating the gain in reactivity for various fusion reaction by using $\kappa$-distributions {\it en lieu}
of Boltzmann ones. To be fair in the comparison, we will compare $\kappa$-distributions with effective
temperature $T_{\mathrm{U}}$ to the Boltzmann distributions with the corresponding core temperature
$T_{\mathrm{core}}$ as defined above.
We have used parameterized cross-sections for D-T and the two channels of the D-D fusion process from \cite{Bosch}.
As customary for fusion processes, the cross-section is described in terms of the $S$-factor capturing the
nuclear physics of the fusion process and softly dependent on the energy (modulo possible resonance phenomena)
and a factor which incorporates the tunneling process through the Coulombian barrier, typically determined
through a Wentzel-Kramers-Brillouin (WKB) approximation

\begin{equation}
\sigma(E)=\frac{S(E)}{E} \exp{\left(-\frac{B_G}{\sqrt{E}}\right)}. 
\label{crosssection}
\end{equation}  
where $B_G$ is the Gamow constant.
Here, the $S$-factor is fitted with a Pad\'e polynomial, following the notation introduced in \cite{Bosch}, that is

\begin{equation}
S(E)=\frac{A1+E(A_2+E(A_3+E(A_4+E A_5))))}{1+E(B_1+E(B_2+E(B_3+EB_4)))}.
\end{equation}
The numerical values of $A_i$ and $B_j$  are determined with a best fit and tabulated in Table IV in \cite{Bosch}.
The corresponding values of the cross-sections are reliable
within few $\%$ in the 3-400 keV energy range, enough for our discussion. Fusion reactivities are
then evaluated by averaging the product of the fusion cross-section and the relative velocities
$v$ between the two nuclei over the corresponding energy distribution.
For Boltzmann-distributed energies, this implies

\begin{equation}
  \langle \sigma v \rangle= \left(\frac{8}{\pi m_r} \right)^{1/2} \frac{1}{(k_BT_{\mathrm{core}})^{3/2}}\int_0^{+\infty}
  dE E e^{-\beta E} \sigma(E),
\end{equation}  
where $m_r$ is the reduced mass of the system made of two colliding nuclei and $\beta$ the inverse temperature $\beta=(k_B T)^{-1}$. 
The reactivity depends on temperature, which allows for a comparison to the parametrized fusion reactivities 
tabulated in Table VII of \cite{Bosch}. In the case of the $\kappa$-distribution, we have

\begin{eqnarray}
\langle \sigma v \rangle &=&\left(\frac{8}{\pi m_r}\right)^{1/2}\frac{C(\kappa)}{(k_BT_{\mathrm{U}})^{3/2}} \times \nonumber \\ 
& & \int_0^{+\infty} dE \frac{E \sigma(E)}{{\left[1+\frac{E}{(\kappa-3/2)k_B T_U}\right]^{\kappa+1}}}.
\end{eqnarray} 
In Fig. 1 we show the dependence on temperature of the reactivities evaluated with averages over Boltzmann
and $\kappa$-distributions for both D-D and D-T fusion reactions. There is a significant advantage in
using $\kappa$-distributions, which should be also beneficial for the D-T reactors at relatively low
temperatures, with reactivity gains of almost two orders of magnitude for temperatures in the few KeV range. 
Fig. 1 also shows that the $\kappa$-distribution is not consistently preferable to the Boltzmann distribution 
over the entire temperature range. The $\kappa$-distributions are peaked at an energy lower than the corresponding 
Boltzmann distribution, with a larger peak probability (see for instance Fig. 3 in \cite{Nicholls}). 
It is therefore understandable that, for a given dependence of the fusion cross-section on energy and with 
respect to the corresponding Boltzmann distribution, they can prevail at lower temperatures, then have an intermediate 
region of marginal or no gain, and prevail again at higher temperatures.
It is also worth to note that the D-D fusion reaction has a Coulomb barrier of about 450 keV.
This means that the high temperature behavior of the reactivity presented in Fig. 1a, purely based on tunneling phenomena, 
is a sort of lower bound on its actual value. Ongoing work here mainly consists in developing kinetic models capable, for 
instance given the power level of neutral beam injection heating, of finding if the plasma energy may be expressed
in terms of a $\kappa$-distribution, including predictions for the value of $\kappa$ and its dependence upon the
experimental parameters. Characterization of non-Boltzmann energy distributions is of current experimental interest,
as witnessed by  recent work at the ASDEX upgrade tokamak \cite{Salewski}. 

\section{Enhancing the reactivity via vacuum polarization effects}

Another possible mechanism which can be used to enhance reactivities consists in exploiting vacuum polarization effects.
The Boltzmann-averaged reactivity of the D-T fusion process is maximum at about 70 keV with a relatively broad peak,
therefore there is no gain in increasing the temperature of a D-T plasma above this value.
As a matter of fact, plasma temperatures of operating D-T tokamaks are in the 10-30 keV range at most.
For D-D plasmas it is instead advantageous to operate at higher temperatures, as the reactivity increases
monotonically until it reaches an even broader peak at temperatures of about 1 MeV.
For instance, the same reactivity of D-T fusion at the temperature of 10 keV is achieved
for D-D fusion only at the temperature of about 300 keV.
Pursuing this high-temperature scenario obviously opens up challenging technological issues due to the
heat irradiated on the first wall, and to the large energy losses due to electron bremmstrahlung.
At the same time, it is interesting to notice that the range of temperatures required for sustained D-D fusion
reactions is not dissimilar from the one in which real electron-positron pairs can be generated.
These electron-positron pairs make the medium in between the two nuclei endowed with an effective
dielectric constant, and therefore in principle may reduce the Coulomb barrier.

The production of particle-antiparticle pairs is a process studied in finite temperature and density quantum field
theory, with applications to a variety of research contexts ranging from astrophysics and cosmology to the observation
of quark-gluon plasma at proton colliders (see for instance \cite{Elmfors,Danielsson,Kapusta}). 
The process is nonperturbative in character, and the particle-antiparticle production rate scales with the mass 
of the particle and the environmental temperature as $\exp(-2m c^2/k_B T)$. Therefore, aiming for electrical polarizability 
effects, and considering the temperature achievable in thermonuclear fusion, we initially limit the attention 
to electron-positron ($e^+e^-$) pairs alone. Even in this case, if one considers plasma temperatures of the order of
100 and 200 keV for instance, the corresponding exponential suppression factors are respectively $3.6 \times 10^{-5}$
and $6.4 \times 10^{-3}$. A detailed discussion of the effective potential between two charges separated by a sea of electrons
and positrons at finite temperature and density is available in \cite{Kaminski}. For temperatures low enough with respect to
the production threshold of $2 m_e c^2 \approx$ 1 MeV and negligible chemical potential, an effective
Yukawa potential has been evaluated \cite{Kaminski}

\begin{equation}
V_{\mathrm{eff}}(r)= -\frac{e^2}{4 \pi \epsilon_0 r} \exp{\left(-\frac{r}{\lambda_{\mathrm{eff}}}\right)},
\label{Kaminski}
\end{equation}  
with the Yukawa range

\begin{equation}
\lambda_{\mathrm{eff}}=\frac{\hbar}{2m_e c^2}\left(\frac{\pi m_e c^2}{2 \alpha_{\mathrm{em}}^2 k_B T}\right)^{1/4}
\exp{\left(\frac{m_e c^2}{2k_B T}\right)}.
\label{lambdaeff}
\end{equation}
where $\alpha_{\mathrm{em}}$ is the fine structure constant.

The above can be interpreted as a temperature-dependent Compton wavelength for the electron-positron pairs, or
alternatively as an effective, temperature-dependent, photon mass. This leads to a space-dependent
weakening of the electrostatic repulsion between two nuclei. In Fig. 2 we plot the dependence on
temperature in a range of interest for D-D fusion processes of various relevant length scales, namely
the effective Yukawa range from Eq. (\ref{lambdaeff}), the Debye length of a plasma with densities usually 
achieved in tokamaks, the distance of minimum approach between two nuclei with a kinetic energy 
equal to $k_B T$, $r_T=e^2/(4 \pi \epsilon_0 k_B T)$ and, for comparison, the (constant) distance at 
which strong interactions give rise to fusion processes, of the order of the proton radius $r_p$.
It is worth to notice that around $T \simeq $ 17 keV the screening due to the electron-positron plasma
takes over the usual Debye screening, and becomes dominant at higher temperatures.

\begin{figure}[t]
\includegraphics[width=1.00\columnwidth, clip=true]{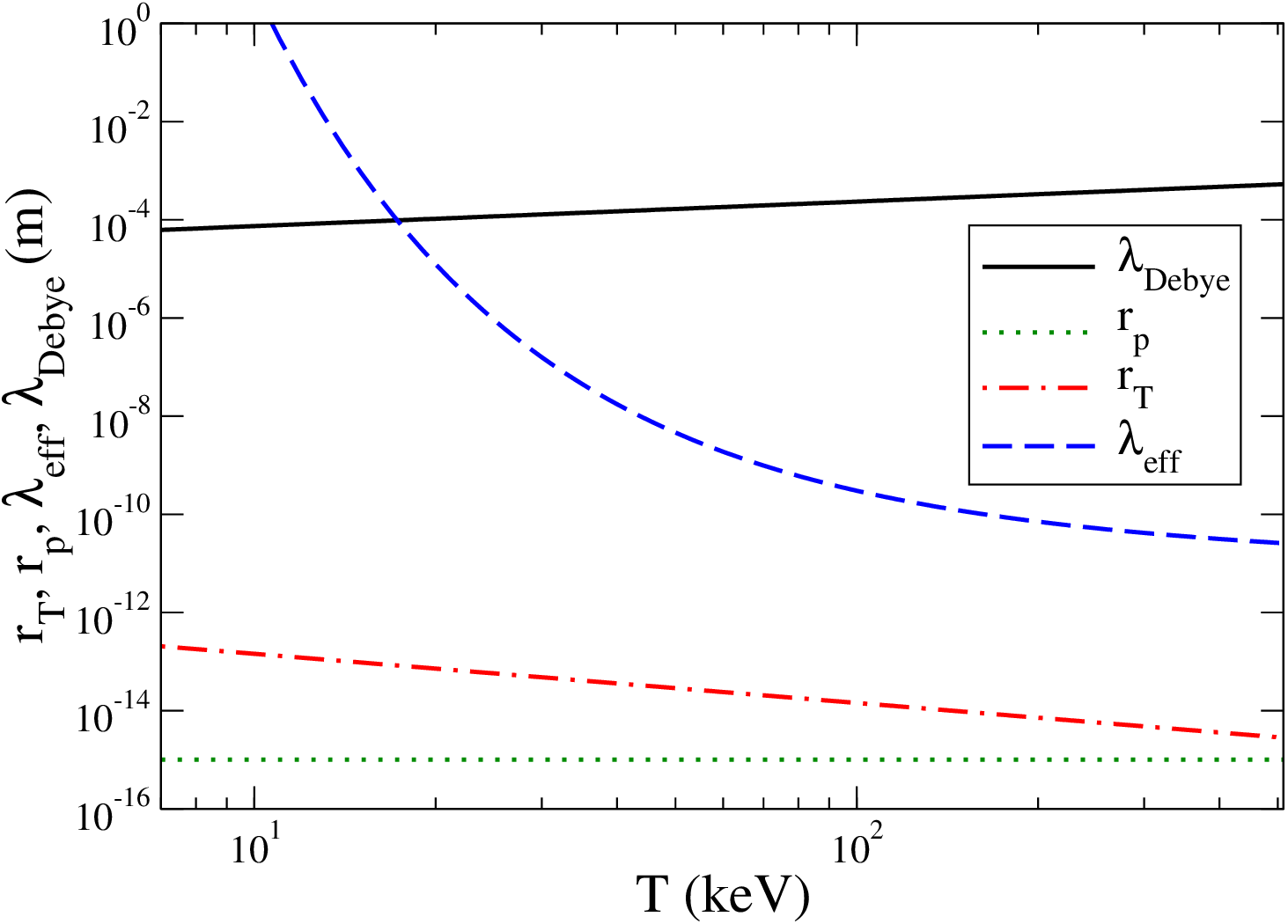}
\caption{Length scales relevant for screening the Coulomb repulsion between nuclei. The effective
Compton wavelength for a finite temperature $e^+e^-$ plasma is evaluated according to Eq. (\ref{Kaminski}).
The Debye length $\lambda_{\mathrm{Debye}}=(\epsilon_0 k_B T/(n e^2))^{1/2}$ is evaluated for a typical ion
density of $n=10^{20} \mathrm{ions/m}^3$, with the consequent crossover between the screening length 
scales $\lambda_{\mathrm{Debye}}$ and $\lambda_{\mathrm{eff}}$ occuring in the 15-20 keV temperature range.}
\label{Fig2}
\end{figure}

Unfortunately, in spite of the favorable scaling of the effective Yukawa range with temperature with
respect to the Debye length, the polarization effect is rather small. This is already visible in Fig. 2 as the 
effective Compton wavelength, although decreasing by increasing temperature at variance with the Debye length, 
is still about four orders of magnitude larger than the average distance of minimum approach between the nuclei.
This is confirmed by a more quantitative analysis by introducing an average polarizability coefficient $\epsilon_r$
such that 

\begin{equation}
\epsilon_r^{-1}= \frac{3}{r_T^3-r_p^3} \int_{r_p}^{r_T} dr r^2 \exp{\left({-r/\lambda_{\mathrm{eff}}}\right)}. 
\label{epsilonr}
\end{equation}  
such that the effective potential in Eq. (\ref{Kaminski}) is obtained via the substitution
$\epsilon_0 \rightarrow \epsilon_r \epsilon_0 $.
The weakening of the electrostatic repulsion due to quantum vacuum screening expressed by Eq. ({\ref{epsilonr}) 
is considered in the classically forbidden volume used for the WKB evaluation of the tunneling rate, {\it i.e.} between 
the nuclei distance $r_p$ (of the order of a proton radius) and the distance of minimum approach $r_T$.
After integration and series expansion of the exponential term, one gets an effective relative
permittivity by which the Coulomb electrostatic repulsion is weakened as
$\epsilon_r \approx 1+r_T/(4 \lambda_{\mathrm{eff}})$ which is negligible for  $r_T/\lambda_{\mathrm{eff}}<<1$. 

The gain in reactivity evaluated in this way should be considered no more than an estimate, providing an
upper bound to the reactivity enhancement, which can be evaluated for instance using screening corrections
as discussed in \cite{Schatzman,Keller,Salpeter}. However, the more energetic nuclei which are mainly
responsible for fusion have a classical distance of approach much smaller than their average value $r_T$,
thus shielding is not effective in that part of the trajectory. For the same reason, the screening effect expected
for a $\kappa$-distribution is even smaller. On the other hand, the electron-positron plasma has a smaller inertia
than the ions, so it can adapt itself to the new situation creating a time-dependent barrier with an intermediate
effectiveness. A dynamical response model is required to accurately describe this effect.

It seems that the spontaneous creation of electron-positron pairs at finite temperature is not enough 
to improve screening of the Coulombian potential, and alternative and more complex schemes should be 
considered. For instance, the possibility to enhance reactivity via controlled microexplosions as a compression 
stage has been discussed in \cite{Shmatov}. One could think to increase the screening by injecting a large 
amount of electrons to upset the densities of electrons and ions. 
The latter situation has also the advantage of improving sympathetic heating of the ions in case electron cyclotron 
resonance heating is implemented, due to the favorable heat capacity matching. 
Unfortunately it is easy to show that the densities required to have at least an electron-positron pair in between 
the distance of minimum approach of two nuclei are prohibitive. A less impossible idea is to intersect the 
confined plasma with the focused beams of an electron-positron collider having enough energy to generate 
$\mu^+\mu^-$ pairs. By replacing the electron mass with the muon mass in Eq. (\ref{lambdaeff}), and provided the effective 
temperature of the $\mu^+\mu^-$ gas is high enough, one can create a situation in which $\lambda_{\mathrm{eff}} \simeq r_T$. 
This is analogous to muon catalysis, but it is hard to imagine that with few intersection points between the core
of the confined plasma and the $e^+e^-$ collider there will be enough heating power to ignite fusion, unless a 
confinement geometry in which the whole  plasma and the $e^+e^-$ beams coexist is designed. 

\section{Energy conversion by hadronic calorimeters and high-efficiency photovoltaic cells}

Since a D-D reactor does not need a lithium blanket, this allows for revising the conversion of the neutron
energy into electrical energy. In particular, the traditional thermal cycles have an overall efficiency limited
to about 30 $\%$. Recent developments in scintillating materials, decades-long experience with hadronic calorimeters,
and progress in photovoltaic conversion may allow for an alternative scheme bypassing the thermal cycle while
achieving comparable efficiency.
The idea is to mimic the solar energy collection. In the latter case, the Sun is emitting with
a peak around the visible region of the electromagnetic spectrum, and the few eV of the photons in this regions
are adequate to create electron-hole pairs in semiconductors.
In our case, since fusion in reactors must occur at higher temperatures to reach reasonable amounts of power, 
a wavelength shifter of the produced photons is required. Therefore, neutrons emitted from half of the D-D fusion 
reactions are progressively absorbed in a calorimeter producing ionization and scintillation. 
A material scintillating with a light yield of 30$\%$ has been recently discovered \cite{Hawrami} and third generation 
photovoltaic cells now under development are expected to reach 70$\%$ efficiency \cite{Conibeer}.
Therefore, combining the two technologies, efficiencies of the order of 20 $\%$ seem within reach.
The power of the reactor in this approach seems limited by the radiation damage induced by neutrons in the
calorimeter, with detectors at the Large Hadron Collider at CERN representing a state of the art design in allowed 
neutron flux (estimated to $2 \times 10^{17}$ neutrons/${\mathrm{cm}}^2$ for an integrated luminosity of 
3000 $\mathrm{fb}^{-1}$ \cite{Turner}), and the light yield has been found to decrease with increasing neutron
fluence \cite{Mdhluli}. 

Such a scheme can be feasible, in light of the maximum affordable neutron flux, for compact low and medium
power fusion reactors, with applications to decentralized electricity production in regions requiring low power densities 
(for instance rural regions, and as a complement to intermittent, renewable sources like wind and solar energy),  
and especially in the sector of maritime transport, with the prospect of a virtually unlimited range of the vessels.
Container vessels, due to their emissions and sheer number, significantly contribute to air and water pollution \cite{Wan}. 
Notice that, due to the compact design of a fusion reactor, the usual request for low-cost per unit of surface of
a photovoltaic cell is not a priority as in extensive solar power plants. This could lead to the search for higher
efficiency--higher price photovoltaic cells, still convenient especially if many (now unaccounted for) environmental
costs are considered.

\section{Conclusions}

We have discussed somewhat futuristic approaches to nuclear fusion both in enhancing
reactivity of D-D fusion processes, and in delivering electricity without the use of a thermal cycle.
It is possible that various aspects of this proposal may become more realistic with dedicated efforts,
as no fundamental technological hurdle seems in sight. The production and control of $\kappa$-distributed
energy for the plasma requires a kinetic approach related, for instance, to the dynamics of neutral 
beam injection heating. Exploiting quantum vacuum effects from existing $e^+e^-$ pairs at temperatures 
one order of magnitude higher than usually achieved in tokamaks, or from intersecting $e^+e^-$ beams 
with the plasma to produce $\mu^+\mu^-$ pairs for efficient Coulomb shielding, will require a radical revision of the 
design of current fusion experiments.  Calorimetry and photovoltaic panels need to be integrated into the unique goal
of energy conversion to bypass the limitations in efficiency intrinsic to the thermal cycle, opening up a vast spectrum 
of applications for fusion energy through compact, low-power plants.

This project could be developed in parallel to the existing ITER-DEMO plan on D-T fusion, with no
subtraction of human and financial resources to these existing and already planned facilities.
By now, the margins of safety for embracing a sustainable, continuous, controllable and clean source
of energy before irreversible climate changes are extremely limited. As such, the ITER-DEMO
project and similar ones under development need to be developed at the maximum speed and extent.

Apart from obviously requesting more resources to the society at large, we believe that extra-resources on parallel designs
as the one sketched here could be available once a consistent part of the scientific community will acknowledge 
where the most urgent, impactful priorities are. In this regard, it is reasonable to expect that the currently
disproportionate emphasis on quantum information technologies -- mainly driven by military and finance secrecy issues
-- will fade away, especially once researchers will recognize the large ratio between promised and delivered
results \cite{Svozil}. A balanced transfer of human and financial resources from quantum information
technology to research in controlled plasma fusion will be highly beneficial for our unique and
irreplaceable Planet Earth.

\acknowledgments
It is a special honor to contribute to the Festschrift for colleague and friend Lev Petrovich Pitaevskii, whose outstanding 
contributions to some of the themes discussed in this note, plasma physics, quantum field theory, and statistical physics, 
will continue to be inspirational for many generations to come.

\end{document}